

\def\simle{^<_{\sim}}
\def\simge{^>_{\sim}}
\magnification = 1200
\baselineskip 14pt plus 2pt
\bigskip

\centerline{\bf ENHANCED HEAVY-ELEMENT FORMATION}
\centerline{\bf IN BARYON-INHOMOGENEOUS BIG-BANG MODELS}
\vskip 0.9in
\centerline{K. Jedamzik and G.M. Fuller}
{\it
\centerline{Physics Department}
\centerline{University of California, San Diego}
\centerline{La Jolla, CA 92093-0319}
}
\centerline{ }
\centerline{G.J. Mathews}
{\it
\centerline{University of California}
\centerline{Lawrence Livermore National Laboratory}
\centerline{Livermore, CA 94550}}

\centerline{ }
\centerline{and}
\centerline{ }
\centerline{T. Kajino}
{\it
\centerline{Physics Department}
\centerline{Tokyo Metropolitan University}
\centerline{Tokyo, Japan}
\centerline{ }
}
\centerline{\bf ABSTRACT}
We show that primordial nucleosynthesis in baryon inhomogeneous
big-bang models can lead to significant heavy-element production
while still
satisfying all of the light-element abundance constraints including
the  low lithium abundance observed in population II stars.
 The parameters which admit
this solution arise naturally from the process of neutrino induced
inflation of baryon inhomogeneities prior to the epoch of
nucleosynthesis.  These solutions entail a small fraction
of baryons ($\simle 2\%$) in very high density regions
with local baryon to photon ratio $\eta^h\ \simeq\ 10^{-4}$, while
most baryons are at
a baryon-to-photon ratio which optimizes the agreement with
light-element abundances.  This model would imply a unique
signature of baryon inhomogeneities in the early universe,
evidenced by the existence of primordial material containing heavy-element
products of proton and alpha-burning reactions with an abundance of $[Z]\ \sim
-6$ to $-4$.
\vskip 0.17in
{\it  Subject headings: early universe - abundances,  nuclear reactions,
nucleosynthesis}\hfill\break

\vfill\eject

\baselineskip 18pt plus 2pt

\centerline{\bf 1. Introduction}

Recently, there has been considerable interest in the
possibility that primordial nucleosynthesis may
have occurred in an environment in which the
baryons were distributed inhomogeneously (Alcock, Fuller, \&
Mathews 1987;  Applegate,  Hogan,  \&   Scherrer  1987; 1988; Fuller,
Mathews, \&  Alcock 1988;
Kurki-Suonio {\it et al.} 1988; 1990;
Malaney \&  Fowler  1988;
Boyd \& Kajino 1989; Terasawa \& Sato 1989abc; 1990; Kajino \&
Boyd 1990;
Kurki-Suonio \& Matzner 1989; 1990; Mathews {\it et al.} 1990, 1993;
Kawano {\it et al.} 1992; Jedamzik, Fuller \& Mathews 1993;
Thomas et al. 1993).
Such studies were at first motivated
by speculation (Witten 1984; Applegate \& Hogan 1985)
that a first-order quark-hadron phase transition at ($T \sim 100$ MeV)
could produce baryon inhomogeneities as baryon number was trapped
within bubbles of shrinking quark-gluon plasma.  Even if this
transition is not first order as some lattice-gauge theory
calculations now seem to imply (e.g. Brown et al. 1990) such
fluctuations might also have been produced by a number of other
processes operating in the early universe
such as late baryogenesis during the electroweak phase
transition (Fuller et al. 1993), a TeV-scale Z(3) QCD
 phase transition (Ignatius, Kajantie, \& Rummukainen 1992), kaon condensation
after
the QCD transition (Nelson 1990), or magnetic
fields from superconducting cosmic strings (Malaney \& Butler 1989).  Other
mechanisms
have also been proposed (see Malaney \& Mathews (1993) for a recent review).

The first calculations (Alcock et al. 1987;  Applegate et al. 1987; 1988;
Malaney \& Fowler 1988; Fuller et al. 1988) of the effects of such
inhomogeneities were particularly interesting since they
seemed to admit a much larger average universal
baryon-to-photon ratio $\eta$ than that allowed
in the standard homogeneous big  bang model
(Wagoner, Fowler,  \&  Hoyle
1967; Schramm \&   Wagoner 1977,  Yang et  al. 1984;
Boesgaard \& Steigman 1985; Olive et al. 1990; Walker et al. 1991; Smith,
Kawano \&
Malaney 1993) while still satisfying the light-element abundance
constraints.  Subsequent calculations
(Kurki-Suonio {\it et al.} 1990;
Malaney \&  Fowler  1988;
Terasawa \& Sato 1989abc; 1990; Kajino \&
Boyd 1990;
Kurki-Suonio \& Matzner 1989; 1990; Mathews {\it et al.} 1990, 1993;
Kawano {\it et al.} 1992; Jedamzik {\it et al.} 1993; Thomas et al. 1993)
 have shown, however, that
when baryon diffusion is properly coupled with
the dynamics of the system, the allowed values of $\eta$ which are
consistent with the light-element abundances are
not significantly different than those of the standard big bang.

Even with this constraint on $\eta$, however, an interest in
baryon inhomogeneities has remained  due to the possibility
that such fluctuations might have produced an observable signature
in the abundances of heavier elements such as
beryllium and boron (Kajino \& Boyd 1990; Malaney \& Fowler
1989; Terasawa \& Sato
1990; Kawano et al. 1991), intermediate mass elements
(Kajino, Mathews \& Fuller 1990)
 or heavy elements (Applegate et al. 1988; Applegate et al.
 1993).  Such possible signatures are
also constrained, however, by the light-element
abundances (e.g. Terasawa \& Sato 1990). In particular,
the population II lithium abundance, Li/H $\sim 10^{-10}$,
constrains the possible abundances of synthesized heavier nuclei
to be quite small (e.g. Alcock et al. 1990; Terasawa \& Sato
1990).

The purpose of this paper, however,
 is to point out a previously unexplored
region of the parameter space for baryon-inhomogeneous big bang
models in which substantial production of heavy-elements is
possible without violating any of the light-element abundance
constraints.  This solution arises naturally from one of the
most likely
scenarios for the production of baryon inhomogeneities in the
first place.  In fact, we discovered this solution
 not by looking for ways to
optimize heavy-element production, but by exploring the
natural evolution of large baryon number-density fluctuations.  If such
fluctuations
are created in an early epoch then they will be damped subsequently by neutrino
induced
heating and expansion to a characteristic baryon-to-photon ratio $\sim
10^{-4}$.
(Jedamzik \& Fuller 1993; Jedamzik et al. 1993).

\centerline{ }
\centerline{\bf 2. Baryon Inhomogeneities and Neutrino Inflation}
\centerline{ }
Let us begin with a discussion of how baryon inhomogeneities
might be generated in the early universe and
how they are most likely to appear.  The most likely scenario
probably involves a first-order phase transition.
It is not clear whether that transition is associated with
the QCD epoch or an earlier time.  Nevertheless,
any process which couples to baryons in a spatially-dependent way
can produce baryon inhomogeneities (Malaney \& Mathews 1993).
For example, in the QCD transition,
inhomogeneities would most likely be produced through a combination of the
limited
permeability of the phase boundary and the higher
thermodynamic solubility of baryon number in the high-temperature
phase (Witten 1984; Fuller et al 1988; Kurki-Suonio 1988).
The same might be
true if inhomogeneities are produced by a kaon condensate (Nelson 1990) after
the QCD transition.

If baryogenesis is associated
with a first-order phase transition at the electroweak epoch
(Kuzmin, Rubakov, \& Shaposhnikov 1985; Cohen, Kaplan \& Nelson 1990;
Dine {\it et al.} 1991; Turok \& Zadrozny 1991; McLerran {\it et al.} 1992)
then it may well be that the net baryon number is generated in a inhomogeneous
fashion (Fuller et al. 1993).
The presence of cosmic strings might also induce baryon inhomogeneities
through electromagnetic (Malaney \& Butler 1989) or  gravitational
interactions.

It is a common misconception that the most natural amplitude
for baryon-number fluctuations is just that given by the
thermodynamic ratio of equilibrium baryon densities in the
two phases of a QCD transition. This as a number $\sim$ 100 for the QCD
transition
(Alcock et al. 1987).
This value for the fluctuation amplitude, however, is unlikely.
It would occur only if complete
equilibrium were maintained in both phases till near the end
of the phase transition followed by a sudden complete drop
from equilibrium.  This would require efficient mixing
of baryon number in both phases and efficient transport of baryon number
across the phase boundary till just near the end of
the phase transition.  Near the end of the phase transition,
the transport of baryon number across the phase boundary
can not be efficient enough to establish chemical equilibrium.
This is because the velocity of the phase boundary continuously increases.
Efficient mixing of baryon number within the two phases will also be unlikely
due to the short mean-free path for baryons
for T $\ge$ 50 MeV.  Furthermore if baryon mixing were this
efficient throughout the phase transition, it would
likely be sufficiently efficient after the phase transition to
homogenize any baryon inhomogeneities long before
the epoch of nucleosynthesis.

Mechanisms which take into account the efficiency for
baryon transport, in particular, can lead to significant baryon
inhomogeneities, and
even in the most conservative conditions (e.g. Kurki-Suonio 1988),
can  produce a central baryon density far in excess
of the thermodynamic equilibrium density (Fuller et al. 1988).
If the baryon fluctuations are produced by such a mechanism,
there is almost no limit as to how high the central baryon-to-photon
ratio can become.  The possibility
of such a high baryon density in fact motivated  Witten (1984) to suggest
this mechanism as a means to form stable quark-matter agglomerates during the
QCD phase transition if such matter were the ground state of
baryonic matter at high density.

In any event, it is certainly possible that baryon inhomogeneities produced
by any of the above mechanisms could reach
a central baryon-to-photon ratio of
$\eta \ge\ 10^{-4}$.  This being the case, it is important to
note the effect of neutrinos on any such large-amplitude
baryon fluctuation (Hogan 1978; Hogan 1988; Heckler \& Hogan 1993; Jedamzik \&
Fuller 1993).
These over dense regions will have a higher pressure contribution from
baryons than their surroundings.
Hence, in order to maintain pressure equilibrium with the surrounding
photon bath, any region of baryon overdensity will be at a slightly lower
temperature than the baryon depleted regions.   However,
since neutrinos have a longer mean free path than photons
they will penetrate into and heat the high-baryon-density regions
causing them to expand and diminish the excess baryon density.

Several recent papers have explored this process
in detail (Heckler \& Hogan  1993; Jedamzik \& Fuller 1993; Jedamzik et al.
1993).  Jedamzik \& Fuller (1993) found that this neutrino-induced damping
of inhomogeneities is more pronounced as the density increases,
such that any fluctuation with baryon-to-photon ratio
in the high-density regions, $\eta^h\ \simge\ 10^{-4}$, is inflated
to $\eta^h\ \approx\ 10^{-4}$.  This is  independent of its initial baryon
density
and holds true over a wide range of initial fluctuation radii.
Fluctuations with $\eta$ less than this are essentially unchanged.

This is illustrated in figure 1 which shows the final
baryon-to-photon ratio $\eta_f$ after neutrino inflation
as a function of initial comoving
radius of a fluctuation.  Lines are drawn corresponding to
different initial baryon-to-photon ratio, $\eta_i$.
The degree to which the fluctuations are damped is related to the
time scale during which
the neutrino mean free path is of order the radius of the fluctuation.  For
larger
fluctuation radii, this time scale becomes shorter due to the rapid increase in
neutrino mean free path as the temperature decreases.  Hence, if the initial
radius of the
fluctuation is too large, neutrinos can not sufficiently damp the fluctuations
to bring them to the limiting value of $\eta\ \sim\ 10^{-4}$.

The effects of neutrino inflation on very-large amplitude baryon
fluctuations with initial baryon-to-photon ratios of order unity
($\eta_i\sim 1$) are not shown in figure 1.
Fluctuations at such high baryon densities could correspond to
stable or metastable low-entropy remnants like strange-quark matter
nuggets (Witten 1984). In this case heating by
neutrinos at temperatures of $T\approx 50$ MeV will cause the
evaporation of such nuggets rather than inflation. Only the very largest
nuggets
will survive complete evaporation (Alcock \& Farhi 1985;
Madsen, Heiselberg \&  Riisager 1986).
Evaporated nuggets can leave behind a fluctuation
with a baryon-to-photon ratio significantly
larger than $\eta\approx 10^{-4}$ even after neutrino inflation.
The exact baryon distribution after evaporation will
depend on the initial baryon
number of the nuggets as well as on bulk properties of quark matter
and strong interaction physics at the surface of the nuggets.

In any case, fluctuations with small initial radius and initial
baryon-to-photon ratio
less than unity will be inflated to a natural amplitude
of $\eta^h\approx 10^{-4}$ by the time of primordial nucleosynthesis.
The average baryon-to-photon value in the baryon-depleted
regions $\eta^l$ (or equivalently the total average baryon-to-photon ratio,
$\eta$)
and the total fraction
of baryons in the high density regions, $f_b$, can then be fixed by the model
for
forming the fluctuations or treated
as free parameters to be fixed by the
constraints from light-element nucleosynthesis.  As noted above, models for the
formation of inhomogeneities are consistent with small values of $f_b$.
We will now show that a small value for $f_b$ is also
desired for agreement with light-element nucleosynthesis.

\centerline{ }
\centerline{\bf 3. Results}
\centerline{ }
Fixing $\eta$ at $10^{-4}$ in the high-density regions
can avoid the overproduction
of $^7$Li.  This  was apparent already in the standard big bang calculations
of Wagoner et al. (1967).  This value for $\eta$
 is, however, much larger than has been considered in most standard
or inhomogeneous big bang models since that early work.  We can
illustrate the effects of a high $\eta^h$ value by first
considering the simplest  possible baryon
inhomogeneous model, that in which the fluctuations are so widely
separated that no significant baryon
diffusion occurs before or during the epoch of primordial
nucleosynthesis.  In this limit the high-density and low-density
regions can be treated (Wagoner et al. 1967; Wagoner 1973; Schramm \& Wagoner
1977;
Yang et al. 1984) as separate standard homogeneous big bang models
with different local values for $\eta$. The nucleosynthesis yields from
these regions can then be averaged after the epoch of nucleosynthesis.
(We will also consider below the effects of baryon diffusion on small-scale
fluctuations.)

The results of standard big bang nucleosynthesis (with three light neutrino
species) at such
high densities are summarized in figure 2 where we have extended the
big-bang nuclear reaction network through mass $A$ = 28.
The point which has
been overlooked in recent studies of baryon inhomogeneous
models is that, for $\eta$ greater than a few times 10$^{-5}$, $^7$Li
is destroyed by the $^{7}$Li($\alpha,\gamma)^{11}$B reaction.  The $^{11}$B
thus formed is then rapidly consumed by a sequence of proton and alpha captures
up to heavy nuclei reminiscent of an $rp$-process (Wallace \&  Woosley 1981).
In fact,
for $\eta \ge\ 10^{-4}$, almost all of the material with atomic mass heavier
than $^4$He is processed to nuclei beyond the end of
our network at $A$ = 28.  Thus, we avoid the  overproduction of $^7$Li
usually associated with averaging over $\eta$ values near the lithium
minimum at $\eta \sim\ 2 \times 10^{-10}$ (e.g. Alcock et al 1987).

It is, of course, still necessary to produce some deuterium, and in so doing,
some
$^3$He and $^7$Li will also be produced.
There is also the problem that the $^4$He abundance is significantly
overproduced, i.e. $Y_p \simeq 0.36$ for $\eta \simeq 10^{-4}$.  The
simultaneous
solution to these constraints fixes allowed values of $\eta^l$ and the
fraction of baryons in the high-density region, $f_b\ =\ f_v(\eta^h/\eta$)
where $f_v$ is the fraction of total volume occupied by the
high density regions,
and $\eta$ is the average baryon-to-photon ratio.

We adopt constraints on observed light-element abundances
from Walker et al. (1991) and Smith et al. (1993) such that
in this simple model with no diffusion, the set of constraint
equations become,
$$0.22\ \le\ f_b Y_h + (1-f_b)Y_l\ \le\ 0.24  \eqno(1a)$$
$$D/H ={X_D^h f_b  + (1-f_b)X_D^l \over
2[X_H^h f_b  + (1-f_b)X_H^l]}\
 \simeq\ (1 - f_b) (D/H)^l\  \ge\ 1.8 \times 10^{-5}  \eqno(1b)$$
$$[D + ^3He]/H\ \simeq\ (1 - f_b) ([D + ^3He]/H)^l\ \le\ 1.0 \times 10^{-4}
\eqno(1c)$$
$$0.8 \times 10^{-10}\ \le\ Li/H\ \simeq\ (1-f_b) (Li/H)^l\ \le\ 2.3 \times
10^{-10}\ \ ,  \eqno(1d)$$
where $X_i$ is the mass fraction of species, $i$, and the
superscripts $h$ and $l$ refer to quantities computed in the
high and low baryon density regions, respectively.  The approximate relations
in (1b)-(1d) are valid for small $f_b$ and insignificant D, $^3$He or $^7$Li
production
in the high baryon density regions.

Figure 3 shows allowed regions of the parameter space of
$\eta^l$, (in units of 10$^{-10}$) as a function of $f_b$.
Clearly there is
a region of $f_b$ and $\eta^l$ for which all of the light-element
abundance constraints can be satisfied.  This region is mostly defined
by the upper limits to helium and [D + $^3$He]/H.  This figure does not include
the extension of the allowed parameter space due to nuclear-reaction
uncertainties, as is
usually done in standard big-bang studies (e.g. Walker et al. 1991;
Smith et al. 1993).  For example, if we had allowed for uncertainties in
helium and deuterium production, the range of allowed $\eta^l_{10}$ would
expand to
$2.8\ \le\ \eta^l_{10}\ \le\ 4.0$  for small $f_b$,  This is just
the usually-quoted standard big-bang limit for 3 light neutrino species.
The most stringent constraint on $f_b$ is from the the upper limit
to the primordial helium abundance which necessitates that at most
only $\sim 2\%$ of the baryons can be in high density regions.  If the
uncertainty
in the neutron half life is also included, this upper limit increases to
as high as $f_b\ \simle\ 0.1$.

It is easy to see from an inspection of
Figure 2 why this region of the parameter space is
allowed.  By placing almost all of the baryons ($\simge 98\%$)
in a low density region for which $\eta^l$ is close to the optimum
standard big bang value, a good fit to all of the light-element
abundances is guaranteed.

So we see that in this large-separation limit for inhomogeneous
models, all of the light-element constraints can be satisfied.
Even though this constraint requires that only a small fraction of the baryons
be in high-density regions, we still find a potentially
interesting (though not necessarily practically observable)
signature from inhomogeneous nucleosynthesis.
For $\eta^h\ \simeq 10^{-4}$ and $f_b \sim 0.01$,
the mass fraction $Z$  of heavy ($A \ge 28$)
or CNO nuclei is $Z  \sim  3 \times 10^{-9}$ ($[Z] \simeq -6.5$)
after averaging with the low-density regions.

The mass fraction in nuclei beyond the end of our network at $A$ = 28 could
reach as high as $[Z] \sim -4$ for $\eta^h \sim 10^{-3}$ and $f_b \sim 0.01$.
A baryon-to-photon ratio as high as $\eta^h\sim
10^{-3}$ after the epoch of neutrino inflation requires an initially
large-amplitude fluctuation of large radius, or equivalently a large
baryonic mass $M_b\sim 5\times 10^{-11}M_{\odot}$ (cf. Fig.1).
Another possibility to obtain a local baryon-to-photon ratio
during primordial nucleosynthesis in excess of $\eta^h\sim 10^{-4}$
is from the evaporation of low-entropy nuggets.
The effects of strange-quark matter
nuggets on primordial nucleosynthesis have been considered
(Schaeffer, Delbourgo-Salvador \&  Audouze 1985; Madsen \&
Riisager 1985), but the evaporation of nuggets before primordial
nucleosynthesis and the possibility of significant heavy element production
during primordial nucleosynthesis has not been addressed before.

It is interesting
to note that a small fraction of baryons initially within
large-mass fluctuations or low-entropy nuggets
might produce a mass fraction in
heavy elements during primordial nucleosynthesis,
in excess of
$[Z] \sim -4$. Thus, observation (e.g. Beers, Preston, \& Shectman 1992)
of very low-metallicity objects
($[Z] \simle -4$) can potentially constrain the existence of
large-mass fluctuations or
low-entropy remnants in the early universe.  The existence of a floor in
primordial
heavy-element abundances could even provide a nucleosynthesis signature of
the physical properties of the fluctuations.

In any case, fluctuations of moderate amplitude and size will be
damped by neutrino inflation to a baryon-to-photon ratio of
$\eta^h\sim 10^{-4}$.
Hence, $[Z] \sim -6$ is probably a good characteristic
abundance for the sum of heavy-element abundances produced by
inhomogeneous nucleosynthesis.

In this case, a possible abundance diagnostic signature includes
C/O and N/O ratios (Kajino  et al. 1990).  A high C/O or N/O ratio
is contrary to what one might expect from
a first generation of massive stars, but does appear for big bang models
with $10^{-5}$ $\simle$ $\eta^h$ $\simle$ $10^{-4}$.  For $\eta^h$ $\simge$
$10^{-4}$,
one should see an anomalously low [O/Fe] ratio, assuming that the
$A \ge 28$ nuclei are burned to Fe.  We realize, of course, that
detecting abundances this low is extremely difficult and
probably not possible with existing
techniques.  Nevertheless, it may eventually be possible to accurately measure
abundances this low, in which case the presence or absence of baryon
inhomogeneities
might be explorable.
\vskip 0.15in
\centerline{{\bf 3.1 Effects of Baryon Diffusion}}
\vskip 0.15in
The results shown on figure 3, of course, correspond to only one possible model
in which the separation between fluctuations is so large that
baryon  diffusion is insignificant.
We have also explored models in which neutron diffusion has been
included but for a reaction network limited to $A \le 12$.
We have considered models in which the proper   (at $T$ = 100 MeV)
separation distance was varied between the QCD horizon (10$^4$ m)
and a small fraction thereof (0.1 m) for fluctuations with fixed $\eta^h$ and
$\eta^l$.
Alternatively, we have considered effects of
baryon diffusion as a function of baryon mass in the fluctuation, $M_b$,
and fraction $f_b$ of the baryons in the high density regions for fixed
$\eta^l$.

Conventionally a regular lattice of fluctuations is described by
four variables, the mean
separation between adjacent fluctuations, $r$, as well as three more
variables.  These can be chosen as
the volume fraction in the high-density region, $f_v$,
the ratio of densities in high-density region and low-density
region, $R$ , and the average baryon-to-photon ratio, $\eta$ (Fuller et al.
1988).  In the present analysis of high-amplitude fluctuations, which include
only a small fraction of the total baryons ($f_b<<1$), it is convenient to
characterize the
fluctuation by baryon mass
instead of the mean separation distance.
Varying $f_b$ at fixed $M_b$ is equivalent to
varying the separation distance.  Varying $M_b$ at fixed $f_b$ corresponds
to a self-similar variation of fluctuation radius and separation distance.

A conversion between the mean separation
and baryon mass is easily obtained, from
$$r\approx 10 {\rm m} \biggl({M_b\over 10^{-21}M_{\odot}}\biggr)^{1\over 3}
\biggl({\eta^l\over 2.8\times 10^{-10}}\biggr)^{-{1\over 3}}
\biggl({f_b\over 0.01}\biggr)^{-{1\over 3}}\ , \eqno(2)$$
where $r$ is the proper mean separation between fluctuations
(at $T=100$ MeV),
and $M_b$ is the baryon mass within a high-density fluctuation.
In equation (2) and what follows it is straightforward to convert
from proper length
at $T_0 = 100$ MeV to proper length at temperature $T$.
To a good approximation, one can do this by simply
assuming that the scale factor $a \propto T_0/T$
for all temperatures $T > 0.1$ MeV.
The baryon mass is related to the
initial (final) fluctuation baryon-to-photon ratio, $\eta_i$
($\eta_f$), and
fluctuation radius, $R_i$ ($R_f$), as follows
$$M_b\approx 4\times 10^{-13}M_{\odot}\eta_i^hR_i^3=4
\times 10^{-13}M_{\odot}
\eta_f^hR_f^3\ ,\eqno(3)$$
where the proper ($T = 100$ MeV) fluctuation radius $R$ is in meters.
Note that the product of baryon-to-photon ratio times the cube of
the fluctuation
radius is not changed by neutrino inflation
because baryon number within the fluctuation is conserved.

With the help of equation 3 and figure 1 the reader is able to
determine the mass
of a fluctuation, $M_b$, its
baryon-to-photon ratio after neutrino inflation,
$\eta_f$ , and radius, $R_f$, assuming $\eta_i$ and $R_i$.

In order to identify the mass scales for which baryon diffusion
affects fluctuations
before and during primordial nucleosynthesis, the baryon
diffusion length,
$d$, should be compared to the radius
of the
fluctuation after neutrino inflation, $R_f$. Baryon diffusion in high-density
regions is
limited by neutron-proton scattering (Applegate et al.
1987, Jedamzik \&  Fuller 1993). The proper (referred
to $T_0=100$ MeV) baryon diffusion
length at temperature $T$, i.e. the average distance a neutron diffuses
between high temperatures and temperature $T$ , is given by
$$d(T)\approx 3\times 10^{-2}m\biggl({\eta_f^h\over
10^{-4}}\biggr)^{-{1\over 2}}
\biggl({T\over 500 keV}\biggr)^{-{5\over 4}}\ .\eqno(4)$$
Clearly, the baryon diffusion length is smaller for a larger value of the
baryon-to-photon ratio in the fluctuation.  For a fluctuation of baryon mass,
$M_b$
$\simle 10^{-21} M_\odot$, the baryon diffusion length at $T \sim 500\ keV$
is comparable to the
fluctuation size, and the fluctuations, therefore, vanish by diffusion before
nucleosynthesis begins.

For mass scales $M_b > 10^{-21}\ M_\odot$, for which the fluctuations can
survive to
the epoch of nucleosynthesis, the abundance yields from models which
include baryon diffusion are not that much different from figure 2.
This is because in high-density regions with $\eta^h \sim 10^{-4}$, nuclear
reactions can
begin to synthesize light elements at temperatures as high as
$T\approx 300-400$ keV.  Thus, in the high-density regions free neutrons are
incorporated into helium soon after the freezeout of weak reactions.
As the weak reactions freeze out, neutrons and protons are no longer rapidly
converted from one into another.  Hence, they obtain separate identities.

Because the neutrons are absorbed so quickly after weak freezeout,
there is not much time for neutron diffusion
to produce large variations in the neutron-to-proton ratio.
Even though some free neutrons in the low-density regions
can diffuse back into the
high-density regions (Malaney \& Fowler 1988),
this back diffusion does not significantly affect
the light element abundances when $f_b << 1$.  This is because light elements
are
produced in low-density regions.  This influx of neutrons into the high-density
regions may somewhat affect the yields of heavy elements which are produced
there.
For the most part, however, the neutron diffusion length in the
high-baryon-density
regions is short and the final abundance yields
should not differ much from that obtained by
simply averaging over regions with different $\eta$ and
fixed $n/p$ ratio as in figure 2.

The results of numerical studies including the effects of baryon diffusion
are presented in figures 4 and 5.
These figures illustrate that
fluctuations more widely separated than a proper distance of
$r=100$ m (at $T = 100$ MeV) or
of a mass scale $\ge 10^{-18}\ M_\odot$
produce nucleosynthesis yields which
 are indistinguishable from those of figure 3 (Mathews et al. 1990).
Similarly, models in which the proper separation distance
(at $T = 100$ MeV)
is $r\simle 5 m$ ($M_b \le 10^{-21} M_\odot$)
are indistinguishable from a homogeneous standard
big bang model with the same average $\eta$.  The reason
is that baryon diffusion prior to nucleosynthesis damps out
the inhomogeneities (Mathews et al. 1990).  The maximum effects
on the nucleosynthesis yields occur for proper separation distances
(at $T = 100$ MeV)
between 10 and 60 m where the separation roughly equals
the neutron diffusion length during primordial nucleosynthesis.
(Applegate et al. 1988; Mathews et al. 1990; Terasawa \& Sato 1990).

Figure 4 identifies the region in the $M_b$ (total baryon mass of the
fluctuation) versus
$f_b$ plane in which heavy-element nucleosynthesis may imply a signature of
high-amplitude
fluctuations.  To produce this figure we used the inhomogeneous nucleosynthesis
code of Jedamzik et al. (1993) which includes hydrodynamic and radiation
transport effects.
We evolved baryon-number fluctuations between $T = 100$ MeV and the end of
primordial
nucleosynthesis at $T \simeq 5$ keV.  The effects of neutrino inflation at high
temperatures
are included in the numerical treatment.

The total baryon mass
within the horizon for the electroweak transition ($M_{EW}$) and the
QCD transition ($M_{QCD}$) is also identified on figure 4.
For $M_b$ $\simge 10^{-18}$ (of order the electroweak
baryon mass scale) and $f_b\ $ $\simle 0.02$
heavy elements are formed  without overproducing lithium or helium.
For less massive fluctuations, $M_b\ \sim\ 10^{-21}$, it is also possible
to satisfy the lithium and helium constraints, but in this case baryon
diffusion erases the inhomogeneities before the onset of primordial
nucleosynthesis
so that the results are indistinguishable from the homogeneous case.

The effects of diffusion on the yields of light abundances
is not very large even at optimum separation distances or masses.  This is
because the high-density regions are constrained to contain only a small
fraction
of the total baryons.  The main contribution to the light-element abundances
stems from the low-density regions which are not much affected by diffusion
from the
high-density regions.

The effects of baryon diffusion on the various
light-element abundances are briefly summarized as follows:

\centerline{\it $Y_p$}
\item{ } Independent of $\eta^h$, the helium mass
fraction decreases by $\Delta Y_p\
\simeq \ 0.0005$ as the separation is decreased from 100 to 6 m.  [For the
remainder of this section all quoted length scales are proper lengths measured
at $T = 100$ MeV.]
As the separation
is further decreased from 6 to 3 m,
the helium mass fraction decreases by  $\Delta Y_p\ \simeq\ 0.002$.
This change is small compared to the
uncertainty in the upper limit to $Y_p$.

\centerline{ }
\centerline{\it D/H, (D+$^3$He)/H}
\item{ } Because these light elements are made
almost exclusively in the low-density regions which are
largely uneffected by baryon diffusion their only dependence
upon the diffusion length is to decrease by about 4\% as
the diffusion length is varied between 50 and 5 m.  Again, this
change is insignificant compared to the uncertainties in the
abundance constraints themselves.

\centerline{ }
\centerline{\it $^7$Li}
\item{ } Lithium is the light-element most significantly
affected by neutron diffusion.  This is largely because of
$^7$Be production at the fluctuation boundary (Mathews et al. 1990).
This causes the
lithium abundance to increase to a maximum of $^7$Li/H $\sim\
2 \times 10^{-10}$ for a separation of 10 m.
For distances less than 6 m or greater than 100 m,
the $^7$Li abundance remains at the value on figure 1 corresponding
 to whatever $\eta^l$ value is chosen.

\centerline{ }

\centerline{\it $^9$Be, $^{11}$B}
\item{ } Since there has been some discussion
of the production of these light nuclei in recent literature,
a mention of their production in the present models should be
made here.  There can be no production of Be in these models,
and almost no $^{11}$B production [X($^{11}$B) $\simle 10^{-18}$] even
with an optimized separation distance (which, like
lithium, can only increase boron by less than a factor of 2).
Therefore, the abundances of these elements can not be a
diagnostic for the kind of inhomogeneities considered here.

\centerline{ }
\centerline{$A \ge 12$}
\item{ } By far the yield which is most sensitive to the
diffusion efficiency is that of the heavy-element abundances.
This sensitivity is illustrated on figure 5.  For separation
distances less than $\simeq$ 6 m the yields are no different
than a standard homogeneous big bang with the same average
value of $\eta$.  In this case most of the yield is
in $^{12}$C at an insignificant abundance relative to hydrogen of less than
10$^{-14}$ (see figure 2).  However, as the diffusion length
is increased to $\simeq$ 100 m, the heavy-element abundances quickly
increases by 5 orders of magnitude due largely to alpha and
proton captures in the high-density regions.
For $\eta^h {\simge} 10^{-4}$, heavy elements are mostly produced in
the form of iron and an anomalously low [O/Fe] would be a diagnostic
abundance signature. For $10^{-5} {\simle} \eta^h {\simle} 10^{-4}$,
anomalously high [C/O] and [N/O] ratios are expected (Kajino et al.
1990). Any
observation suggesting the presence of primordial heavy elements with
abundances
as large as [Z] $\simge$ - 6
could be a strong verification of the presence of baryon
inhomogeneities in the early universe.

\centerline{ }
\centerline{\bf 4. Conclusions}
We have shown that there exists a plausible region of the
parameter space for baryon-inhomogeneous primordial nucleosynthesis
models in which all of the light-element abundance constraints are
satisfied, yet which also leads to the unique production of potentially
observable abundances of heavy nuclei.  Although the abundances of these heavy
nuclei
are small, they can be as much as 6 to 8 orders of magnitude larger than
in the homogeneous big bang.  The region of the
parameter space in which heavy-element abundances are generated
is naturally motivated by models for the formation
of baryon inhomogeneities, for example during a first-order phase transition,
and by numerical simulations of the effect of neutrino-induced
heating and expansion of baryon inhomogeneities prior
to the epoch of primordial nucleosynthesis.  Hence, this
region of the parameter space should be taken seriously.

The conditions most likely to result from the expansion of
a high baryon density region by neutrino
inflation which also satisfy the light-element abundance constraints,
imply that a mass fraction as large as $[Z] \sim -4$ to $-6$ could
be in the form of nuclei with $A > 28$ and/or  CNO nuclei after primordial
nucleosynthesis.
Thus, a search for a lower limit to the abundance ratios
of CNO or heavy nuclei to the light primordial abundances could some day be a
definitive indicator of the presence or absence of large-amplitude
baryon inhomogeneities in the early universe.

\centerline{ }
\centerline{\bf 5. Acknowledgements}

    This work  was supported
in part by the National Science Foundation through
Grants PHY91-21623 and INT88-15999, and through IGPP LLNL Grant
93-22. It was also performed in part under
the auspices of the  U.S. Department of  Energy by the Lawrence
Livermore National Laboratory under contract number W-7405-ENG-48
and DoE Nuclear Theory grant SF-ENG-48.

\centerline{ }
\centerline{\bf References}
\item{ }Alcock, C.R. \& Farhi, E. 1985, Phys. Rev., D32, 1273
\item{ }Alcock, C.R. \& Hogan, C. J., Phys. Rev., D31, 3037
\item{ }Alcock, C.R., Fuller, G.M., \&  Mathews, G.J. 1987, ApJ, 320,  439
\item{ }Alcock, C.R., Dearborn, D.S., Fuller, G.M., Mathews, G.J., \&  Meyer,
    B.S. 1990, Phys. Rev. Lett., 64, 2607
\item{ }Applegate, J.H., Hogan, C.J., \&   Scherrer, R.J. 1987, Phys. Rev.,
D35, 1151
\item{ }Applegate, J.H., Hogan, C.J., \&   Scherrer, R.J. 1988, ApJ, 329, 592
\item{ }Applegate, J.H., Cowan, J. J., Rauscher, T., Thielemann, F.-K., \&
Wiescher, M. 1993,
ApJ, submitted
\item{ }Beers, T. C., Preston, G. W., \& Shectman, S. A. 1992, ApJS 103, 1987
\item{ }Boesgaard,  A.M.  \&  Steigman, G. 1985, ARA\&A, 23, 319
\item{ }Boyd, R.N.\&  Kajino, T. 1989, ApJ, 336, L55.
\item{ }Brown, F. R.  et al. 1990, Phys. Rev. Lett., 65, 2491
\item{ }Cohen, A. G., Kaplan, D. B., \&  Nelson, A. E., 1990,  Phys.  Lett.,
245B, 561
\item{ }Dine, M., Huet, P., Singleton, R., \&  Susskind, L., 1991, Phys.
Lett., 257B, 351
\item{ }Fuller,  G.M.,  Mathews, G.J., \&  Alcock, C.R. 1988, Phys. Rev., D37,
1380
\item{ }Fuller, G. M. et al. 1993, ApJ, submitted
\item{ }Heckler, A. \& Hogan, C. J. 1993 Phys. Rev. D, submitted
\item{ }Hogan, C. 1978, MNRAS, 185, 889
\item{ }Hogan, C. 1988, in {\it Primordial Nucleosynthesis}, W. J. Thompson,
B. W. Carney, \&  H. J. Karwowski, Eds., (Singapore: World Scientific)
\item{ }Ignatius, J., Kajantie, K., \&  Rummukainen, K. 1992, Phys. Rev. Lett.,
68, 737.
\item{ }Jedamzik, K. \& Fuller, G. M. 1993, ApJ submitted
\item{ }Jedamzik, K., Fuller, G. M., \& Mathews, G, J, 1993, ApJ submitted
\item{ }Kajino, T. \&  Boyd, R.N. 1990, ApJ, 359, 267
\item{ }Kajino,  T.,  Mathews,  G.J., \&  Fuller, G.M. 1990, ApJ, 364, 7
\item{ }Kawano, L.H., Fowler, W.A., Kavanagh,  R.W., \& Malaney, R.A., 1991,
ApJ, 372, 1
\item{ }Kurki-Suonio, H. 1988, Phys. Rev. D37, 2104.
\item{  }Kurki-Suonio, H., \&  Matzner, R. A. 1989, Phys. Rev. D39, 1046
\item{  }Kurki-Suonio, H., \&  Matzner, R. A. 1990, Phys. Rev. D42, 1047
\item{ }Kurki-Suonio, H., Matzner, R.A., Centrella, J., Rothman, T., \&
    Wilson, J.R. 1988, Phys. Rev., D38, 1091
\item{ }Kurki-Suonio, H., Matzner, R. A., Olive, K. A., \& Schramm, D. N. 1990,
ApJ, 353, 406
\item{ }Kuzmin, V. A., Rubakov, V. A., \&  Shaposhnikov, M. E., 1985,  Phys.
Lett., 155B, 36
\item{ }Madsen, J., Heiselberg, H., \& Riisager, K. 1986, Phys. Rev., D34, 2947
\item{ }Madsen, J., \& Riisager, K. 1985, Phys. Lett., B158, 208
\item{ }Malaney,  R.A. \&  Butler, M.N. 1989, Phy. Rev. Lett., 62, 117
\item{ }Malaney,  R.A.   \&  Mathews,   G.J. 1993, Phy. Rep., in press
\item{ }Malaney, R.A. \&  Fowler, W.A. 1988, ApJ,  333, 14
\item{ }Malaney, R.A. \&  Fowler, W.A. 1989, ApJ,  354, L5
\item{ }Mathews,  G.J., Alcock, C.R., \&  Fuller, G.M. 1990, ApJ, 349, 449
\item{ }Mathews, G.J., Meyer, B.S., Alcock, C.R., \&  Fuller, G.M. 1990,  ApJ,
358, 36
\item{ }Mathews, G.J., Schramm, D.N., \& Meyer, B.S. 1993, ApJ, 404, 476
\item{ }McLerran, L., Shaposhnikov, N., Turok, N., \& Voloshin, M. 1991, Phys.
Lett., B256, 451
\item{ }Nelson, A. 1990, Phys. Lett., B240, 179
\item{ }Olive, K.A., Schramm, D.N., Steigman, G., \& Walker, T.P. 1990, Phys.
Lett., B236, 454
\item{ }Schaeffer, R. Delbourgo-Salvador, P., \& Audouze, J. 1985, Nature, 370,
407
\item{ }Schramm, D.N. \&  Wagoner, R.V. 1977, Ann. Rev. Nucl. Part. Sci., 27,
37
\item{ }Smith, M. S., Kawano, L. H. \&  Malaney, R. A., 1993, ApJS, 85, 219
\item{  }Terasawa, N., \&  Sato, K. 1989a, Prog. Theor. Phys., 81, 254.
\item{  }Terasawa, N., \&  Sato, K. 1989b, Phys. Rev., D39, 2893.
\item{  }Terasawa, N., \&  Sato, K. 1989c, Prog. Theor. Phys., 81, 1085.
\item{ }Terasawa, N., \&  Sato, K., 1990, Ap. J. Lett., 362, L47.
\item{ }Turok, N. \& Zadrozny, J. 1990, Phys. Rev. Lett., 65, 2331
\item{ }Thomas et al. 1993, ApJ submitted
\item{ }Wagoner,  R.V., Fowler, W.A., \&  Hoyle, F. 1967, ApJ, 148, 3
\item{ }Wagoner, R. V. 1973, ApJ, 197, 343
\item{ }Walker, T.P., Steigman, G., Schramm, D.N., Olive, K.A., \& Kang, H.
1991, ApJ, 376, 51
\item{ }Wallace, R. K. \& Woosley, S. E. 1981, ApJS, 45, 389
\item{ }Witten, E. 1984, Phys. Rev. D30, 272
\item{ }Yang, J., Turner, M.S., Steigman, G., Schramm, D.N., \& Olive, K.A.
1984, ApJ, 281, 493

\vfill\eject

\centerline{\bf Figure Captions}

\centerline{ }
\item{Figure 1}Final baryon-to-photon ratio, $\eta_f$ of neutrino-inflated
fluctuations as a function of initial fluctuation radius (comoving at 100 MeV).
Results are shown for four different initial values of the baryon-to-photon
ratio, $\eta_i$.

\centerline{ }
\item{Figure 2}Nucleosynthesis yields as a function of baryon-to-photon
ratio, $\eta$ for a reaction network extending from light nuclei
to $A$ = 28.  These yields are for locally homogeneous regions
sufficiently separated that baryon diffusion is insignificant.
Note that for $\eta\ \simge\ 10^{-4}$ the yields of nuclei heavier than
helium are almost entirely in massive nuclei with essentially no
production of other light elements.

\centerline{ }
\item{Figure 3.} Light-element abundance constraints on the baryon-to-photon
ratio in the low density regions, $\eta^l$,
as a function of the fraction, $f_b$, of baryons in high-density
regions.
Note that both quantities, $\eta^l$ and $f_b$, are plotted on a
logarithmic scale.
For this model the high baryon density region
is fixed by neutrino
induced inflation of the fluctuations
to be $\eta^h\ =\ 10^{-4}$.

\centerline{ }
\item{Figure 4.} Heavy-element nucleosynthesis signatures of high-amplitude
fluctuations ($\eta_i \ge\ 10^{-4}$, $\eta^l = 2.8 \times 10^{-10}$).
Identified are
regions of the total
baryon mass $M_b$ (in units of $M_\odot$) contained in a single fluctuation
versus the fraction of baryons in these
high-density fluctuations, $f_b$.  Also identified on this figure are the total
baryon masses
within the horizon for the electroweak transition ($M_{EW}$) and the
QCD transition ($M_{QCD}$).  For $M_b\ \simge\ 10^{-18}$ (roughly the
electroweak
baryon mass scale) and $f_b\ \simle 0.02$,
heavy elements are formed   without overproducing lithium or helium.
For less massive fluctuations, $M_b\ \simle\ 10^{-21}$, it is also possible
to satisfy the lithium and helium constraints, but in this case baryon
diffusion erases the inhomogeneities before the onset of primordial
nucleosynthesis
so that the results are indistinguishable from the homogeneous
standard big bang and no heavy-element signature emerges.

\centerline{ }
\item{Figure 5.} An example of the effects of baryon diffusion on heavy-element
abundances (A $\ge$ 12).  Shown are averaged yields for heavy elements
as a function of proper separation distance
in m at $T\ =\ 100$ MeV.  For this figure we have arbitrarily set
$f_b = 0.0001$, $\eta^l = 2.8 \times 10^{-10}$,
and $\eta^h = 1.3 \times 10^{-5}$.  At separations less than 6 m, the
results are indistinguishable from a standard homogeneous big
bang with $\eta \sim 2.8 \times 10^{-10}$ because baryon diffusion before
nucleosynthesis homogenizes the fluctuation.
For separation distances greater than
$\sim 100$ m, the results are essentially identical to figure 2.
A transition between the two limits occurs when the neutron diffusion length
during nucleosynthesis is comparable to the separation distance.

\end